\documentclass[12pt,a4paper]{article}

\usepackage{amscd,amsfonts,amsmath,amssymb,amsthm,bbm,bm,latexsym,mathrsfs}
\usepackage{epsfig,graphics,graphicx,natbib,subfigure,longtable}
\usepackage[english]{babel}
\usepackage[usenames]{color}
\usepackage{bookmark}
\usepackage{booktabs}
\usepackage{sgame}
\usepackage[top=1.7in, bottom=1.7in, left=0.9in, right=0.9in]{geometry}
\allowdisplaybreaks
\makeatother

\theoremstyle{remark}

\theoremstyle{plain}

\begin{document}

\title{\vspace{-50pt} A Note on the Posterior Inference for the Yule--Simon Distribution}%\thanks{Authors research is supported by funding from the European Union, Seventh Framework Programme FP7/2007-2013 under grant agreement SYRTO-SSH-2012-320270, by the Institut Europlace of Finance, ``Systemic Risk grant", the Global Risk Institute in Financial Services, the Louis Bachelier Institute, ``Systemic Risk Research Initiative", and by the Italian Ministry of Education, University and Research (MIUR) PRIN 2010-11 grant MISURA.}}

\author{
\hspace{-20pt} 
Fabrizio Leisen\textsuperscript{a} \hspace{15pt}
 Luca Rossini\textsuperscript{b} \hspace{15pt} Cristiano Villa\textsuperscript{a}\thanks{Email Adresses: \hspace{-2pt}\href{mailto:fabrizio.leisen@gmail.com}{fabrizio.leisen@gmail.com} (F. Leisen); \href{mailto:luca.rossini@unive.it}{luca.rossini@unive.it} (L. Rossini); \href{mailto:cv88@kent.ac.uk}{cv88@kent.ac.uk} (C. Villa).}
        \\ 
        \vspace{5pt}
        \\
        {\centering {\small \textsuperscript{a} 
        University of Kent, UK \hspace{18pt}
        \textsuperscript{b}
            Ca' Foscari University of Venice, Italy}} \vspace{5pt} \\
     }

\date{}
\maketitle

%%%%%%%%%%%%%%%%%%%%%%%%%%%%%%%%%%%%%%%%%%%%%%%%%%%%%%
%%         	Abstract			%%%
%%%%%%%%%%%%%%%%%%%%%%%%%%%%%%%%%%%%%%%%%%%%%%%%%%%%%%
\abstract{The Yule--Simon distribution has been out of the radar of the Bayesian community, so far. In this note, we propose an explicit Gibbs sampling scheme when a Gamma prior is chosen for the shape parameter. The performance of the algorithm is illustrated with simulation studies, including count data regression, and a real data application to text analysis. We compare our proposal to the frequentist counterparts showing better performance of our algorithm when a small sample size is considered. \\

}

%%%%%%%%%%%%%%%%%%%%%%%%%%%%%%%%%%%%%%%%%%%%%%%%%%%%%%
%%    	     	Intro			%%%
%%%%%%%%%%%%%%%%%%%%%%%%%%%%%%%%%%%%%%%%%%%%%%%%%%%%%%
\section{Introduction}
\label{Intro}
The purpose of this work is to show that a Gamma prior on the shape parameter of the Yule-Simon distribution yields to a straightforward Gibbs sampling scheme, allowing for an efficient and effective approach to Bayesian inference. 
The Yule-Simon distribution \citep{Yule25,Simon55} is mainly employed when the center of interest is some sort of frequency in the data. For example, \cite{Gal16} highlight that the heavy-tailed property of the Yule--Simon distribution allows for extreme values even for small sample sizes. In particular, they claim that the above property is suitable to model short survival times which, due to the nature of the problem, happen with relatively high frequency. 
 To the best of our knowledge, the sole Bayesian proposal to deal with the Yule-Simon distribution has been discussed in \cite*{LeiRosVil16}, where the problem is tackled from an objective point of view. 

The algorithm we propose is based on a stochastic representation of the Yule-Simon distribution as a mixture of Geometric distributions. This naturally suggests a data augmentation scheme which can be employed to address Bayesian inference. In particular, the choice of a Gamma prior leads to explicit full conditional distributions.

To illustrate the performance of the above algorithm, we discuss examples of common applications of the Yule--Simon distribution; namely, count data regression and text analysis. In both cases we compare our inferential results to the respective classical counterparts.

The structure of the paper is as follows. In Section \ref{BApp} we present the data augmentation scheme and the consequent algorithm. The proposed method is then illustrated by means of simulations in Section \ref{Simu}, where we consider both a single i.i.d. sample and a count data regression. Section \ref{Real} is reserved to the application of the proposed algorithm to word frequency text analysis. The last Section \ref{Concl} is dedicated to final remarks.

%%%%%%%%%%%%%%%%%%%%%%%%%%%%%%%%%%%%%%%%%%%%%%%%%%%%%%
%%    	     	Bayesian semi-conj			%%%
%%%%%%%%%%%%%%%%%%%%%%%%%%%%%%%%%%%%%%%%%%%%%%%%%%%%%%
\section{Bayesian Inference}
\label{BApp}

The Yule--Simon distribution has the following probability function:
\begin{equation}
f(k;\rho) = \rho\, \mbox{B}(k,\rho+1), \qquad k=1,2,\ldots \mbox{ and } \rho>0, \label{Yule_Org}
\end{equation}
where $\mbox{B}(\cdot,\cdot)$ is the beta function and $\rho$ is the shape parameter. \cite{Yule25} proposed the distribution derived in \eqref{Yule_Org} in the field of biology; in particular, to represent the distribution of species among genera in some higher taxon of biotic organisms.
Later on, \cite{Simon55} noticed that the above distribution can be observed in other phenomena, which appear to have no connection among each others. These include, the distribution of word frequencies in texts, the distribution of authors by number of scientific articles published, the distribution of cities by population and the distribution of incomes by size. 

The probability distribution defined in \eqref{Yule_Org} can be seen as a mixture of Geometric distributions. Precisely, let $W$ be an exponentially distributed random variable with parameter $\rho$, and let $K$ be a Geometric distribution with probability of success equal to $e^{-W}$. Therefore, it is easy to see that the Yule-Simon distribution can be recovered as the marginal of the random vector $(K,W)$, i.e. 
\begin{align}
f(k;\rho)&=\int_0^{\infty} e^{-w}(1-e^{-w})^{k-1} \rho e^{-\rho w} dw .\label{Compo}
\end{align}
The above description of the Yule-Simon distribution is crucial to define a data augmentation scheme in a Bayesian setting.

Suppose to consider the following Bayesian model:
\begin{equation}\label{BayModel}
\begin{split}
k_1,\dots,k_n|\rho&\sim f(k;\rho)\\
\rho&\sim \mbox{Gamma}(a,b),
\end{split}
\end{equation}
where $f(k;\rho)$ is the Yule-Simon distribution defined in \eqref{Yule_Org}. The likelihood function of the above model, conditionally to the parameter $\rho$, is the following:
\begin{align}
L(\textbf{k},\rho)&=\prod_{i=1}^n \int_0^{\infty} e^{-w_i}(1-e^{-w_i})^{k_i-1}\rho e^{-\rho w_i} dw_i \notag \\
&=\int_{(0,\infty)^n} \prod_{i=1}^n e^{-w_i}(1-e^{-w_i})^{k_i-1}\rho e^{-\rho w_i} d\textbf{w} \notag \\
&= \int_{(0,\infty)^n} L(\textbf{k},\textbf{w},\rho) d\textbf{w} \label{lik}
\end{align}
where $\textbf{k}=(k_1,\dots,k_n)$ is a vector of observations, $\textbf{w}=(w_1,\dots,w_n)$ is a vector of auxiliary variables, and
\begin{equation}
L(\textbf{k},\textbf{w},\rho)=\prod_{i=1}^n e^{-w_i}(1-e^{-w_i})^{k_i-1}\rho e^{-\rho w_i} .\label{liklik}
\end{equation}
In order to perform the Bayesian analysis of the model introduced in \eqref{BayModel}, we consider the following augmented version of the posterior distribution:
$$\pi(\rho,\textbf{w}|\textbf{k})\propto L(\textbf{k},\textbf{w},\rho) \pi(\rho),$$
where $\pi(\rho)\propto \rho^{a-1}e^{-b\rho}$ is the Gamma prior. To sample from the posterior distribution we adopt a Gibbs sampling scheme and compute the full conditional distribution. It is straightforward to note that 
\begin{equation}
p(w_i|w_{-i},\textbf{k},\rho) \propto  e^{-\rho w_i} e^{-w_i}(1-e^{-w_i})^{k_i-1}.  \notag
\end{equation}
The change in variable $t_i=e^{- w_i}$, leads to a full-conditional distribution which is distributed as a $\mbox{Beta}(\rho+1,k_i)$. On the other hand, the full-conditional distribution for $\rho$ is
\begin{equation}
p(\rho|\textbf{k},\textbf{w})\propto \rho^{a+n-1} e^{-\rho(b+ \sum_{i=1}^n w_i)} \sim \mbox{Gamma}\left(a+n, b+\sum_{i=1}^n w_i \right). \notag\label{fullR}
\end{equation}
To sum up, the updating rule of the Gibbs sampler is as follows:
\begin{itemize}
\item Sample $t_i|\rho,k_i \sim \mbox{Beta}(\rho+1,k_i)$, for $i=1,\dots,n$;
\item Compute $w_i=-\log{t_i}$, for $i=1,\dots,n$;
\item Sample $\rho|\textbf{w},\textbf{k} \sim \mbox{Gamma}\left(a+n, b+\sum_{i=1}^n w_i \right)$.
\end{itemize}
We show that the performance of the above algorithm on both simulated data (Section \ref{Simu}) and real data (Section \ref{Real}).

%%%%%%%%%%%%%%%%%%%%%%%%%%%%%%%%%%%%%%%%%%%%%%%%%%%%%%
%%    	     	Simmulated Example			%%%
%%%%%%%%%%%%%%%%%%%%%%%%%%%%%%%%%%%%%%%%%%%%%%%%%%%%%%
\section{Simulation Study}
\label{Simu}

In this section we analyse the performance of the proposed algorithm by considering a single i.i.d. sample generated from a Yule--Simon distribution (Section \ref{sc_singleiid}), and on a regression model for count data where the shape parameter of the Yule--Simon distribution is modelled in a similar fashion to the one in the classical Poisson regression (Section \ref{Count}).

\subsection{Single i.i.d. sample}
\label{sc_singleiid}
This section is devoted to test the performance of the data augmentation algorithm on simulated data. To do this, we sample from a Yule--Simon distribution with two values of the parameter, $\rho=0.80$ and $\rho=5.00$. For each value of the parameter, we have simulated samples of different sizes, respectively $n=30$, $n=100$ and $n=500$. Note that the choice of a relatively small sample size has the purpose to leverage on the Bayesian property of giving sensible results even when the information coming from the data is limited.

For the simulations, we have chosen a Gamma prior with shape parameter $a=0.25$ and rate parameter $b=0.05$. The choice was made with the intent of having a large variance in the prior, reflecting a fairly large prior uncertainty. The Gibbs sampler is run for $50,000$ iterations, with a burn-in period of $10,000$ iterations. This is repeated 20 times per sample to capture the variability in the procedure. Table \ref{TabrhoSim} displays the summary statistics of the posteriors, that is, the mean, the median and mean square errors from these two indexes. Both in terms of central value and mean square error the simulation results are excellent, proving the soundness of the algorithm and, more in general, of the whole proposed approach.

\begin{table}[h!]
\centering
\begin{tabular}{ccccccc}
\hline
$\rho$ & $n$ & Mean & Median  & MSE Mean & MSE Median & Fixed-Point Alg\\ %& Std. Dev.
\hline
0.80 & 30 & 0.80 & 0.78 & 0.00002 &  0.00041 & 0.79 \\ % & 0.17 
0.80 & 100 & 0.80 & 0.80  & 0.00160 & 0.00190 & 0.78 \\ % & 0.09
0.80 & 500 & 0.80 & 0.80 & 0.00002 & 0.00001 & 0.80 \\
\hline
5.00 & 30 & 5.00 & 4.56 & 0.00046 & 0.19000 & 4.42\\ % & 2.1 
5.00 & 100 & 4.82 & 4.70 & 0.03600 & 0.10000 & 4.66 \\ % & 1.08 
5.00 & 500 & 4.90 & 4.87  & 0.00990 & 0.01670 & 4.85 \\ %& 0.49
\hline
\end{tabular}
\caption{Summary statistics of the posterior distributions for the parameter $\rho$ of the simulated data from a Yule--Simon distribution with different values of $\rho=\{0.80, 5.00\}$ and sample sizes $n=\{30, 100, 500\}$.}
\label{TabrhoSim}
\end{table}

\begin{figure}[h!]
\centering
	\subfigure[]
  	{\includegraphics[width=6.75cm]{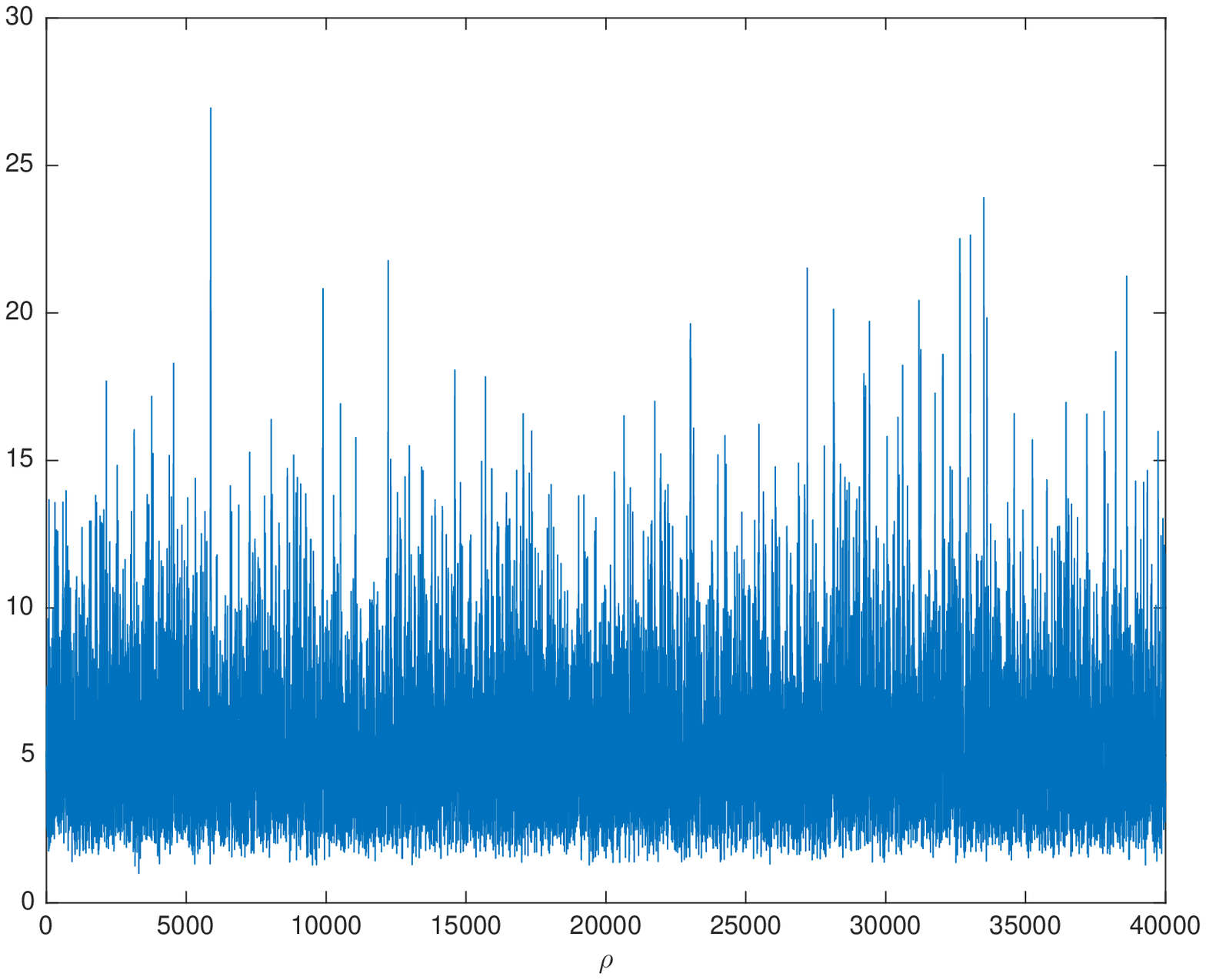}}
  	\subfigure[]
   {\includegraphics[width=6.75cm]{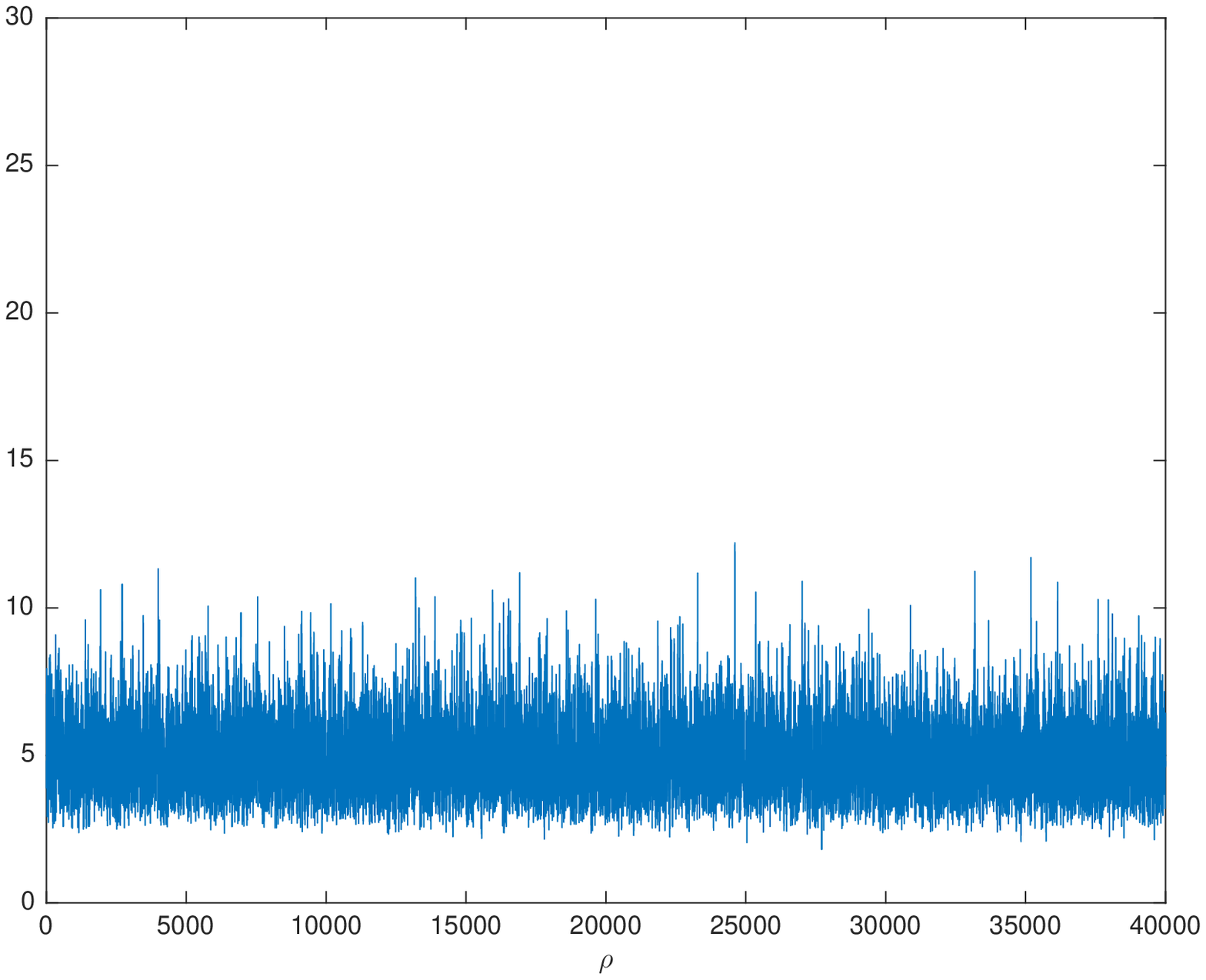}}
   \subfigure[]
   {\includegraphics[width=6.75cm]{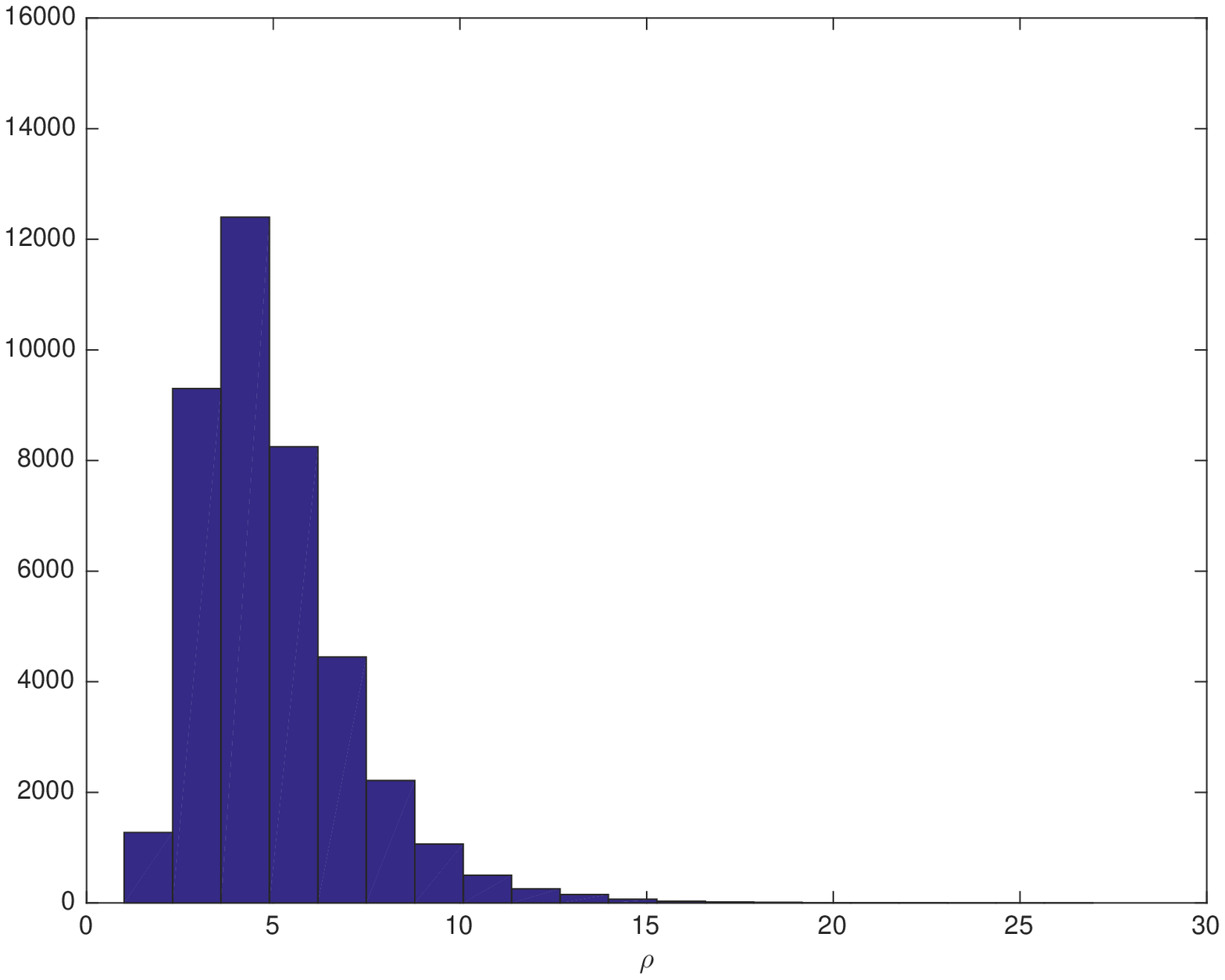}}
   \subfigure[]
   {\includegraphics[width=6.75cm]{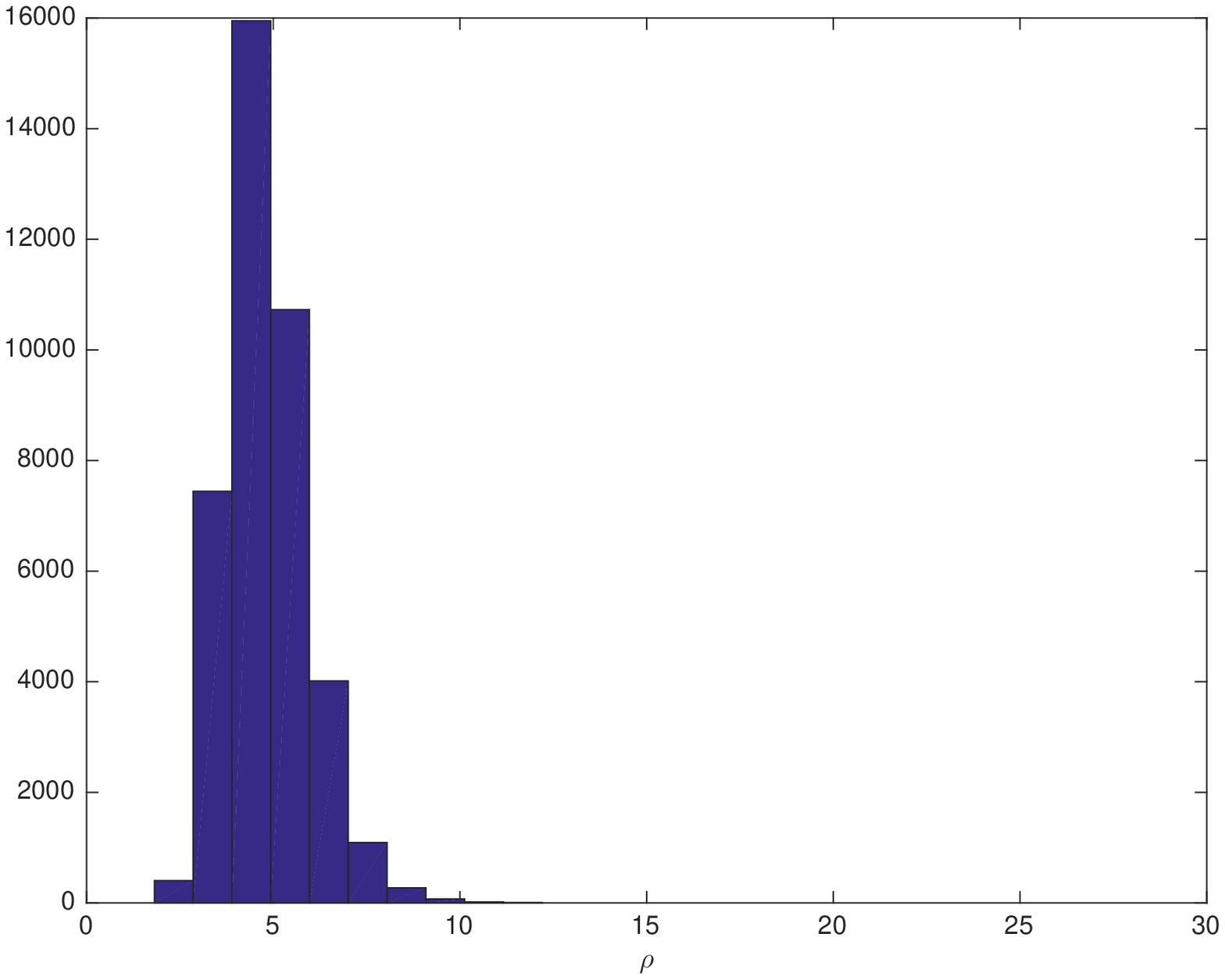}}
   \subfigure[]
   {\includegraphics[width=6.75cm]{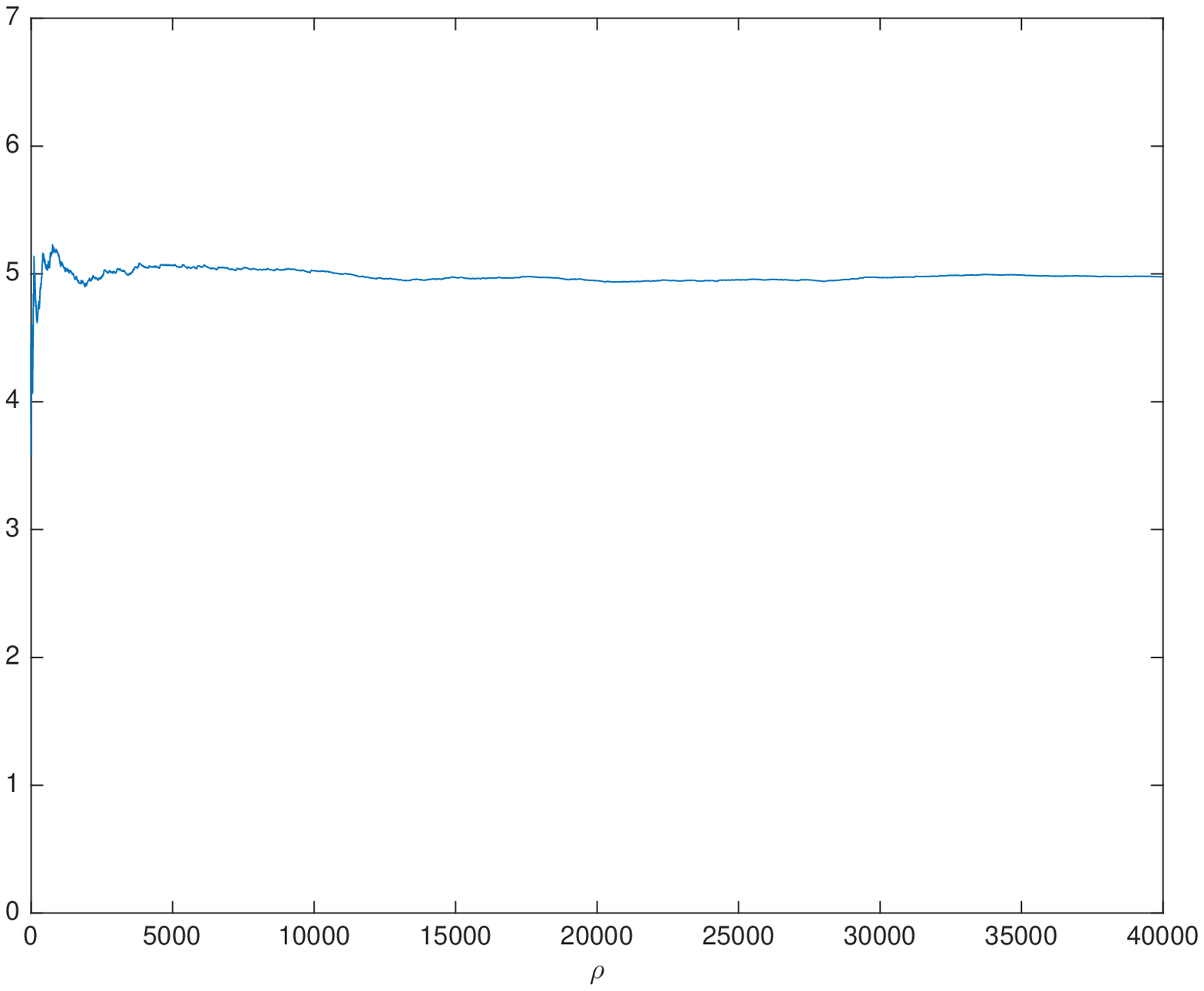}}
   \subfigure[]
   {\includegraphics[width=6.75cm]{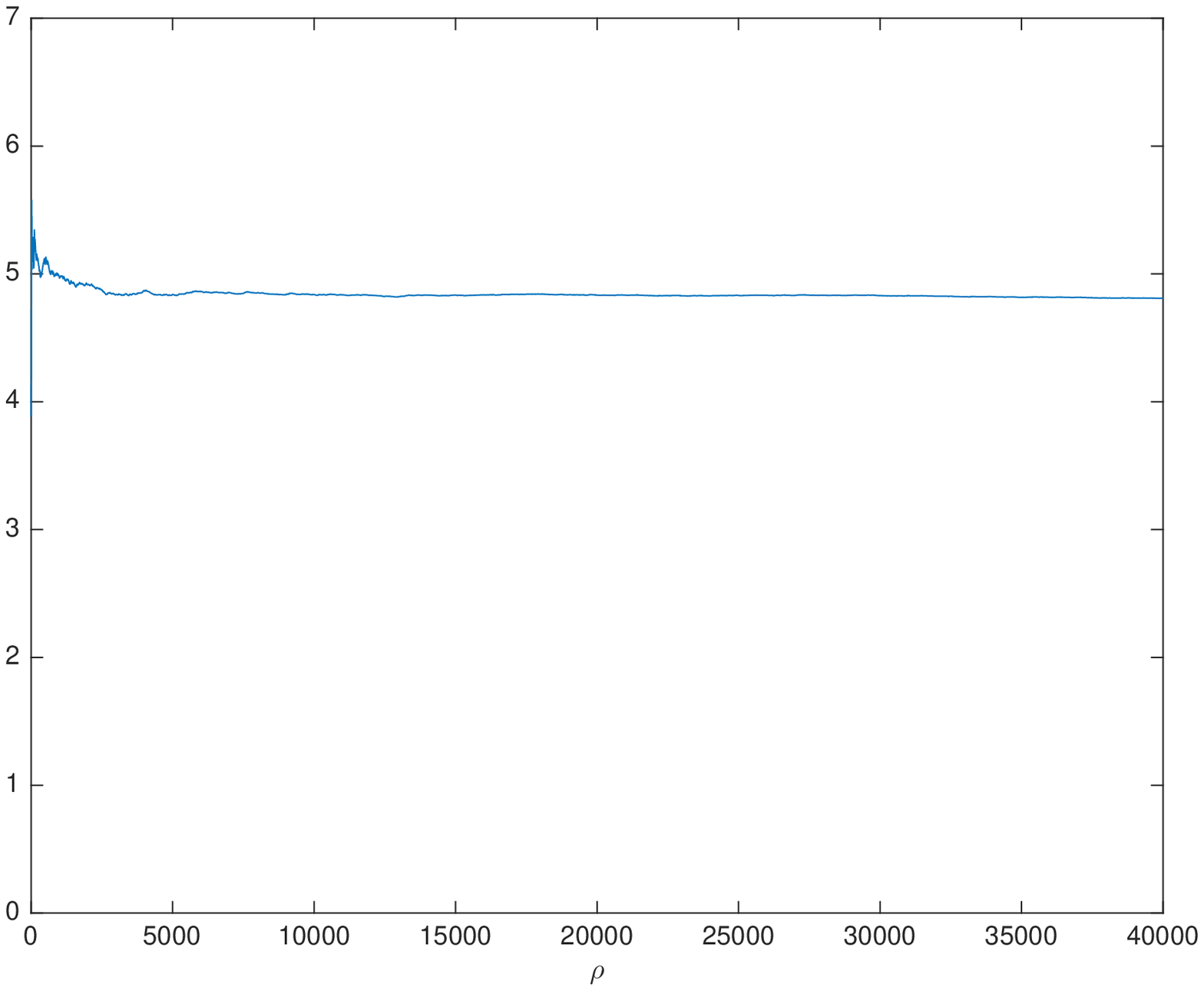}}
\caption{Posterior sample (top), posterior histogram (middle) and progressive mean (bottom) for the simulation study of a Yule--Simon distribution with $\rho=5$ and sample size $n=30$ (left) and $n=100$ (right).}
\label{FigSimRho}
\end{figure}

As an example, in Figure \ref{FigSimRho} we show the posterior results for one simulation of the sample of size $n=30$ from the Yule--Simon with $\rho=5$, and one simulation from the same distribution with $n=100$. We see that the chains exhibit a good mixing and that the means converge to the true values rather quickly. In detail, we have the posterior mean equal to 4.98 for $n=30$ and equal to 4.81 for $n=100$, and the $95\%$ credible intervals are, respectively, $(2.22,10.17)$ and $(3.10,7.30)$. As one would expect, the credible interval for the smaller sample size is larger than the one obtained with $n=100$. This is reflected in the histogram in Figure \ref{FigSimRho} as well.

Although not shown here, we have performed the simulation study on other values of the parameter, ranging from $0.1$ to $10$, obtaining results in line with the above ones.

\subsection{Count data regression}
\label{Count}
In a count data regression model we are interested in the relations between the probability of a dependent variable $k_i$ and the vector of independent variables $x_i$. The model is based on the following three assumptions:
\begin{enumerate}
\item the observation $k_i$ follows the Yule--Simon distribution with parameter $\rho_i$, i.e.
\begin{equation}
f(k_i;\rho_i)=\rho_i B(k_i,\rho_i+1), \quad k_i=1,2,\dots, \quad \rho_i>0; \notag
\end{equation}
\item the parameters of interest are modelled in the following way:
\begin{equation}
\rho_i= \exp{(\textbf{x}_i'\bm{\beta})}, \quad \, i=1,\dots,n, \notag
\end{equation}
where $\bm{\beta}$ is a $(n_{\beta}\times 1)$ vector of parameters and $\textbf{x}_i'=(1,x_{i2},\dots,x_{in_{\beta}})$ is a $(1\times n_{\beta})$ vector of regressors including a constant;
\item the observation pairs $(k_i,x_{i}), i=1,\dots,n$ are independently distributed.
\end{enumerate}
For sake of illustration we focus on the case with one regressor only, although the arguments can easily be extended to include multiple regressors. Therefore, we have $\bm{\beta}'=(\beta_0,\beta_1)$, $\textbf{x}_i'=(1,x_{i2})$ and $\rho_i=\exp{\{\beta_0+\beta_1 x_{i2}\}}$.  Assuming a standard bivariate normal prior for $\bm{\beta}$, we obtain the following augmented version of the posterior distribution:
\begin{align}
\pi(\bm{\beta},\textbf{w},\textbf{x}|\textbf{k})\propto \left[ \prod_{i=1}^n e^{-w_i}(1-e^{-w_i})^{k_i-1}\right] \exp{\left\{\sum_{i=1}^n \textbf{x}_i' \bm{\beta}\right\}} \left[ \prod_{i=1}^n e^{-e^{\textbf{x}_i'\bm{\beta}}w_i}\right]e^{-\frac{1}{2}\bm{\beta}'\bm{\beta}}. \notag
\end{align}
Therefore, the full conditional distribution for the parameter of interest $\beta$ is given by:
\begin{align}
\pi(\bm{\beta}|\textbf{w},\textbf{x},\textbf{y}) &\propto \left[\prod_{i=1}^n \exp{\left\{-e^{\textbf{x}_i' \bm{\beta}}w_i\right\}}\right] \exp{\left\{-\frac{1}{2}\bm{\beta}'\bm{\beta}+\sum_{i=1}^n \textbf{x}_i' \bm{\beta}\right\}}. \label{MH}
%&= \prod_{i=1}^n \exp{\left\{-e^{\textbf{x}_i' \bm{\beta}}w_i\right\}}\mathcal{N}_2\left(\bm{\beta}\Big|\sum_i \textbf{x}_i',I\right)
\end{align}
As the expression in \eqref{MH} is not an explicit known distribution, Monte Carlo methods have to be used. In particular, we adopt a Metropolis within Gibbs to obtain samples from the posterior distribution. We use a random walk proposal and the Gibbs sampler for the count data regression is as follows:
\begin{itemize}
\item Sample $t_i|\beta_0,\beta_1,x_i,k_i \sim \mbox{Beta}\left(\exp{\{\beta_0+\beta_1x_{i2}\}}+1,k_i\right)$, for $i=1,\dots,n$;
\item Compute $w_i=-\log{t_i}$, for $i=1,\dots,n$;
\item Sample $\bm{\beta}|\textbf{w},\textbf{k},\textbf{x}$ from the random walk Metropolis-Hastings algorithm.
\end{itemize}

We test the proposed data augmentation algorithm on two simulated data sets: for the first data set we have $(\beta_0,\beta_1)=(1.5,-1.0)$, and for the second one we have $(\beta_0,\beta_1)=(-0.5,5.0)$. In both cases, the regressor values are sampled from a uniform $(0,1)$. We ran $50,000$ iterations with a burn-in period of $10,000$ iterations, and this has been repeated $20$ times per sample. For comparison purposes, we use the R function (\emph{VGLM}) developed by \citet*{Yee08,VGAM} in the package VGAM. The function allows us to estimate the vector generalized linear model (see \citet*{Yee14,Yee15}), when we consider a Yule--Simon distribution. Table \ref{TabCounSim} shows the posteriors mean, median, mean square errors and credible intervals for the two different scenarios. Overall, the results obtained by applying our algorithm are very close to the true parameter values. As noted in Section \ref{sc_singleiid}, the Bayesian approach outperforms the frequentist for small sample sized.

 In both cases and for all the different sample sizes, the results are interesting for our approach and in particular, as seen in the previous simulated example, for small sample size the results are better from a Bayesian perspective with respect to the frequentist approach.

\begin{table}[h!]
\centering
\begin{tabular}{ccccccc}
\hline
$n$ & $\bm{\beta}$ & Mean & Median  & MSE Mean & $95\%$ C.I. & VGLM \\
\hline
30 & $\beta_0=-0.5$ & -0.5 & -0.5 & 0.0012 & (-0.7,-0.2)  & -0.2 \\
 & $\beta_1=5.0$ & 5.0 & 5.0  & 0.0014 & (4.7,5.2) & 7.7 \\
\hline
30 & $\beta_0=1.5$ & 1.6 & 1.6 & 0.0035 & (1.3,1.8) & 3.0 \\
& $\beta_1=-1.0$ & -1.0 & -1.0 & 0.0025 & (-1.2,-0.7) & -0.9 \\
\hline
100 & $\beta_0=-0.5$ & -0.6 & -0.6 & 0.0069 & (-0.8,-0.4) &-0.7 \\
& $\beta_1=5.0$ & 4.9 & 4.9 & 0.0071 & (4.7,5.2) &4.8 \\
\hline 
100 & $\beta_0=1.5$ & 1.4 & 1.4 & 0.0103 & (1.2,1.6) &1.4 \\
& $\beta_1=-1.0$ & -1.0 & -1.0 & 0.0021 & (-1.3,-0.8) &-1.2\\
\hline
500 & $\beta_0=-0.5$ & -0.5 & -0.5 & 0.0000 & (-0.7,-0.3) & -0.5 \\
& $\beta_1=5.0$ & 5.0 & 5.0 & 0.0029 & (4.7,5.2) & 5.1 \\
\hline
500 & $\beta_0=1.5$ & 1.5 & 1.5 & 0.0002 & (1.3,1.7) & 1.5\\
& $\beta_1=-1.0$ & -1.0 & -1.0 & 0.0004 & (-1.2,-0.8) & -0.9\\ 
\hline
\end{tabular}
\caption{Summary statistics of the posterior distributions for the parameter $(\beta_0,\beta_1)$ of the Yule--Simon regression with $(\beta_0,\beta_1)=\{(-0.5,5.0); (1.5,-1.0)\}$ and sample sizes $n=\{30, 100, 500\}$ and VGLM estimators.}
\label{TabCounSim}
\end{table}

\begin{figure}[h!]
\centering
	\subfigure[]
  	{\includegraphics[width=6.75cm]{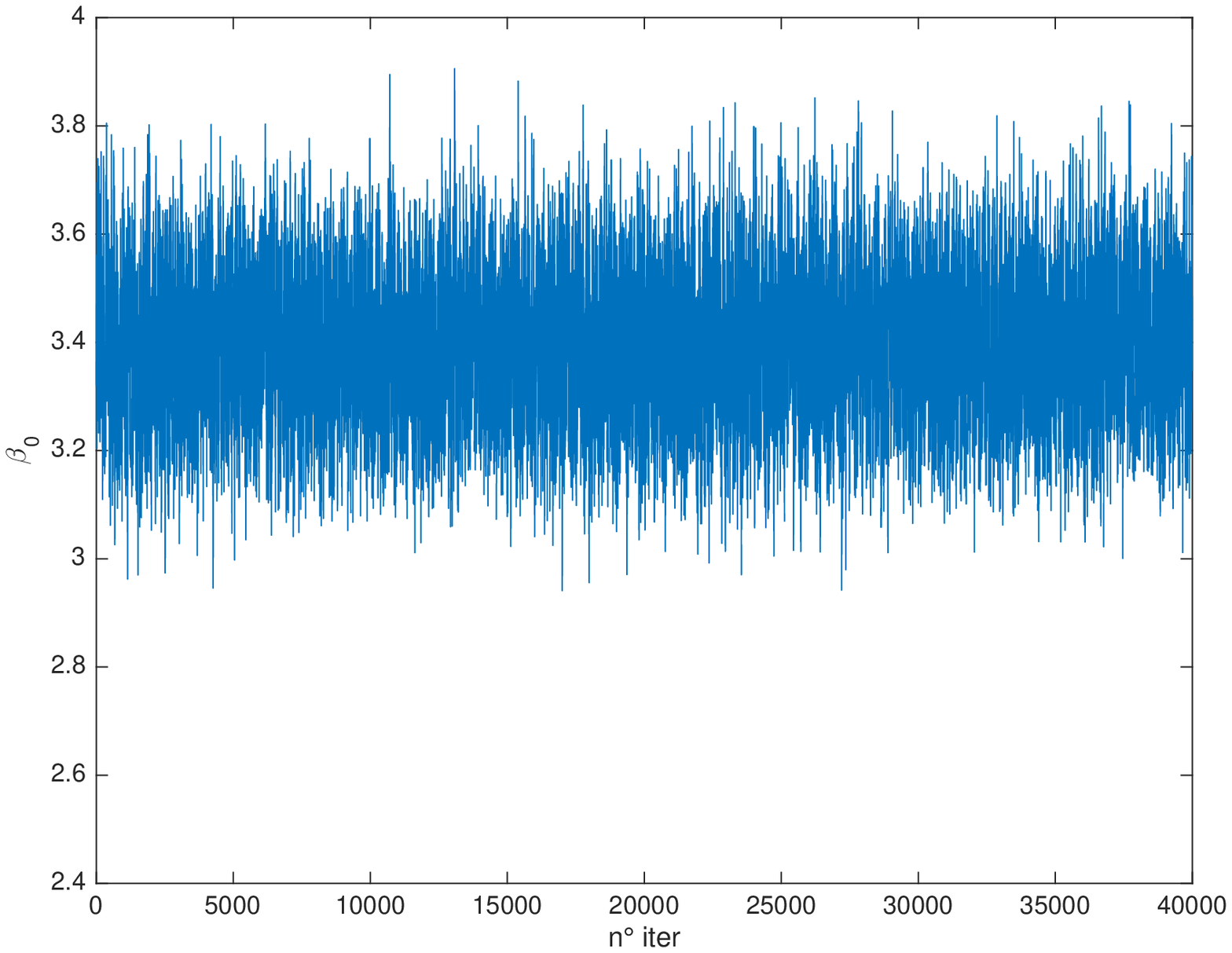}}
  	\subfigure[]
   {\includegraphics[width=6.75cm]{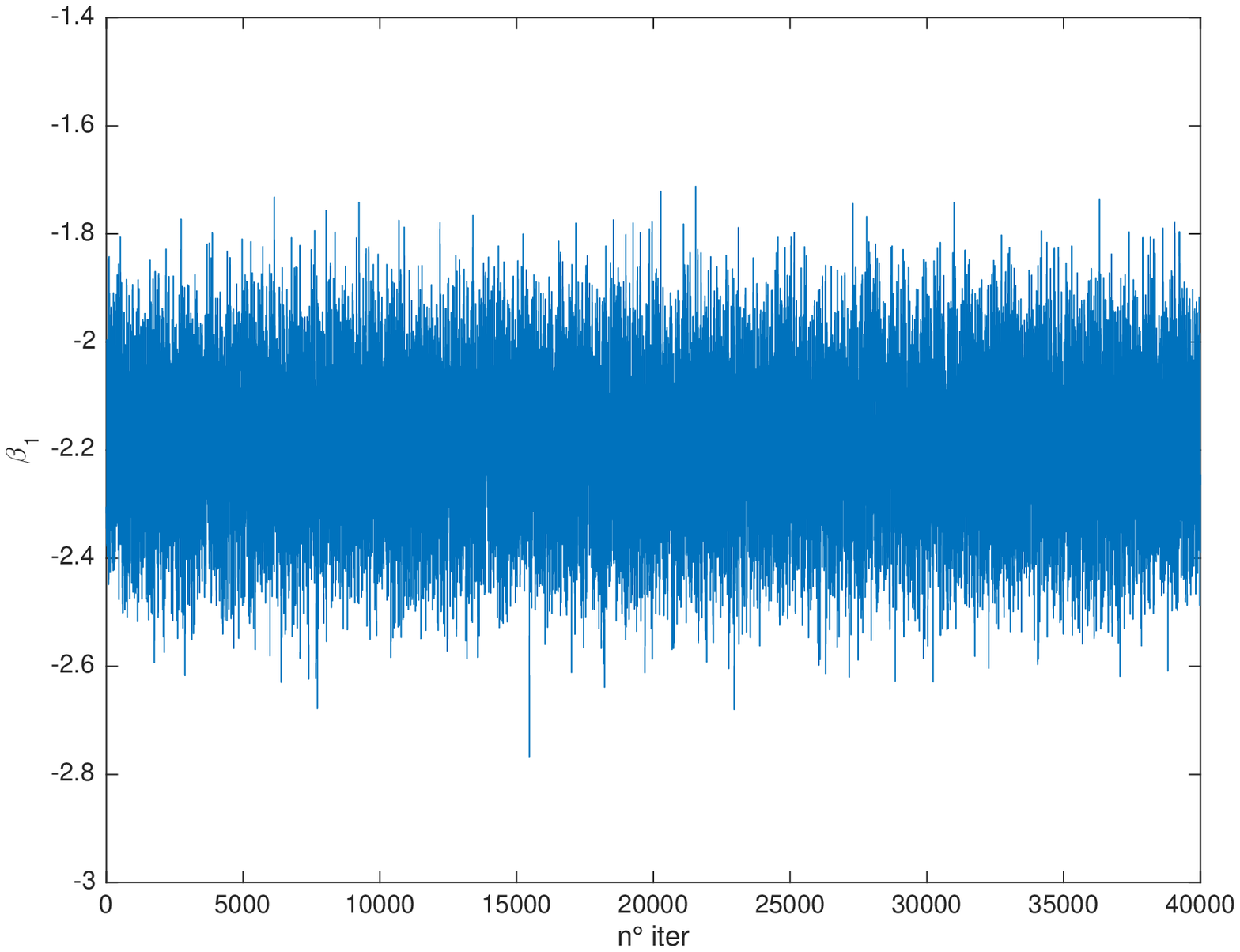}}
   \subfigure[]
   {\includegraphics[width=6.75cm]{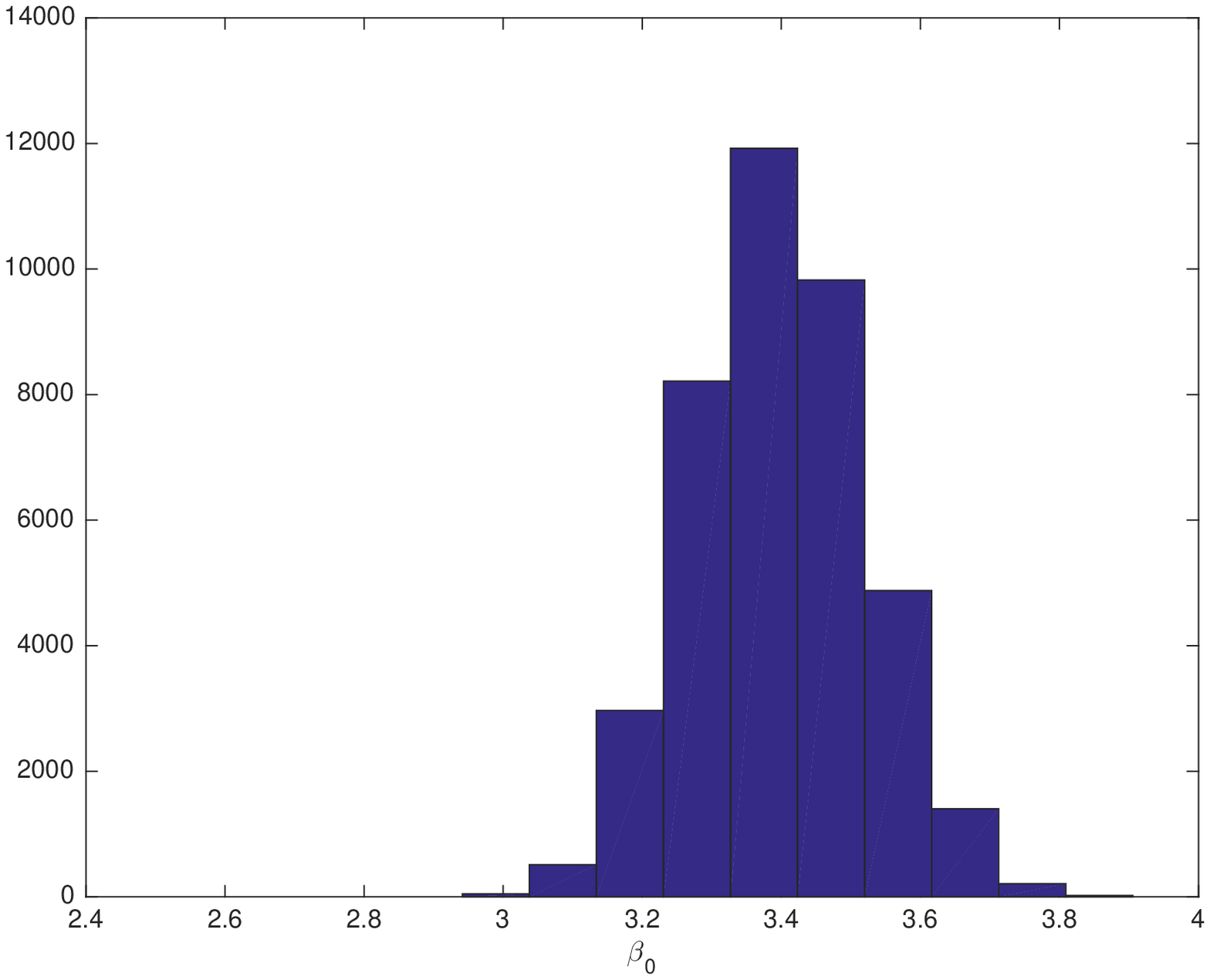}}
   \subfigure[]
   {\includegraphics[width=6.75cm]{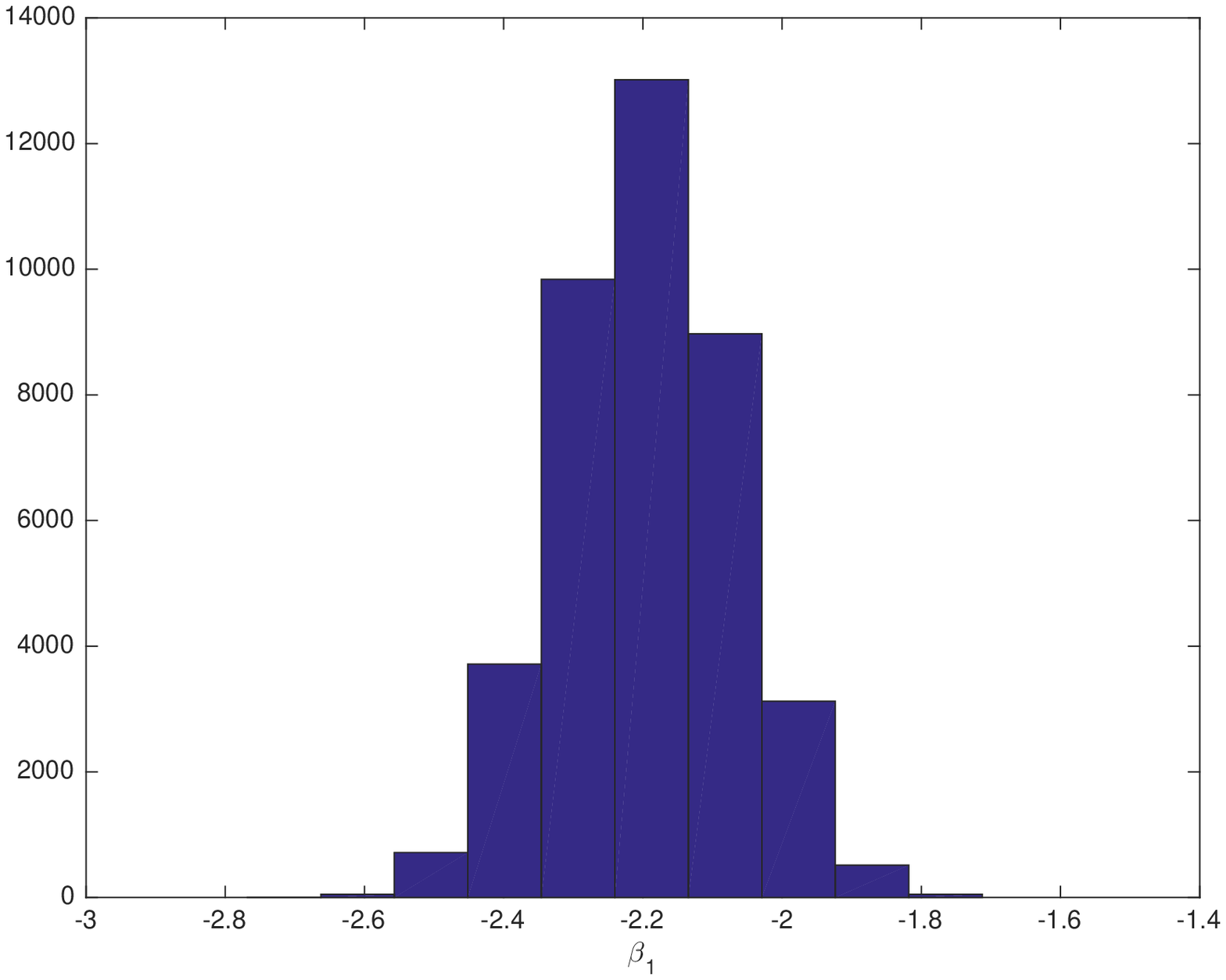}}
\caption{Posterior sample (top) and posterior histogram (bottom) for the simulation study of a count data regression with $\beta_0=3.5$ (left) and $\beta_1=-2.2$ (right) and sample size $n=300$.}
\label{FigSimCount1}
\end{figure}

To better illustrate the performance we have simulated 300 observations for a case with $\beta_0=3.5$ and $\beta_1=-2.2$. Figure \ref{FigSimCount1} shows the posterior samples and the posterior histograms obtained with a Gibbs sampler run for 50,000 iterations with a burn-in period of 10,000. We see that for both parameters of the regression the chain has a good mixing, and the posterior means for $\beta_0$ and $\beta_1$ are, respectively, $3.40$ and $-2.195$. The $95\%$ credible intervals are, respectively, $(3.2,3.6)$ and $(-2.4,-2.2)$ which comfortably contain the true values of the parameters.

\begin{figure}[h!]
\centering
	\subfigure[]
  	{\includegraphics[width=6.75cm]{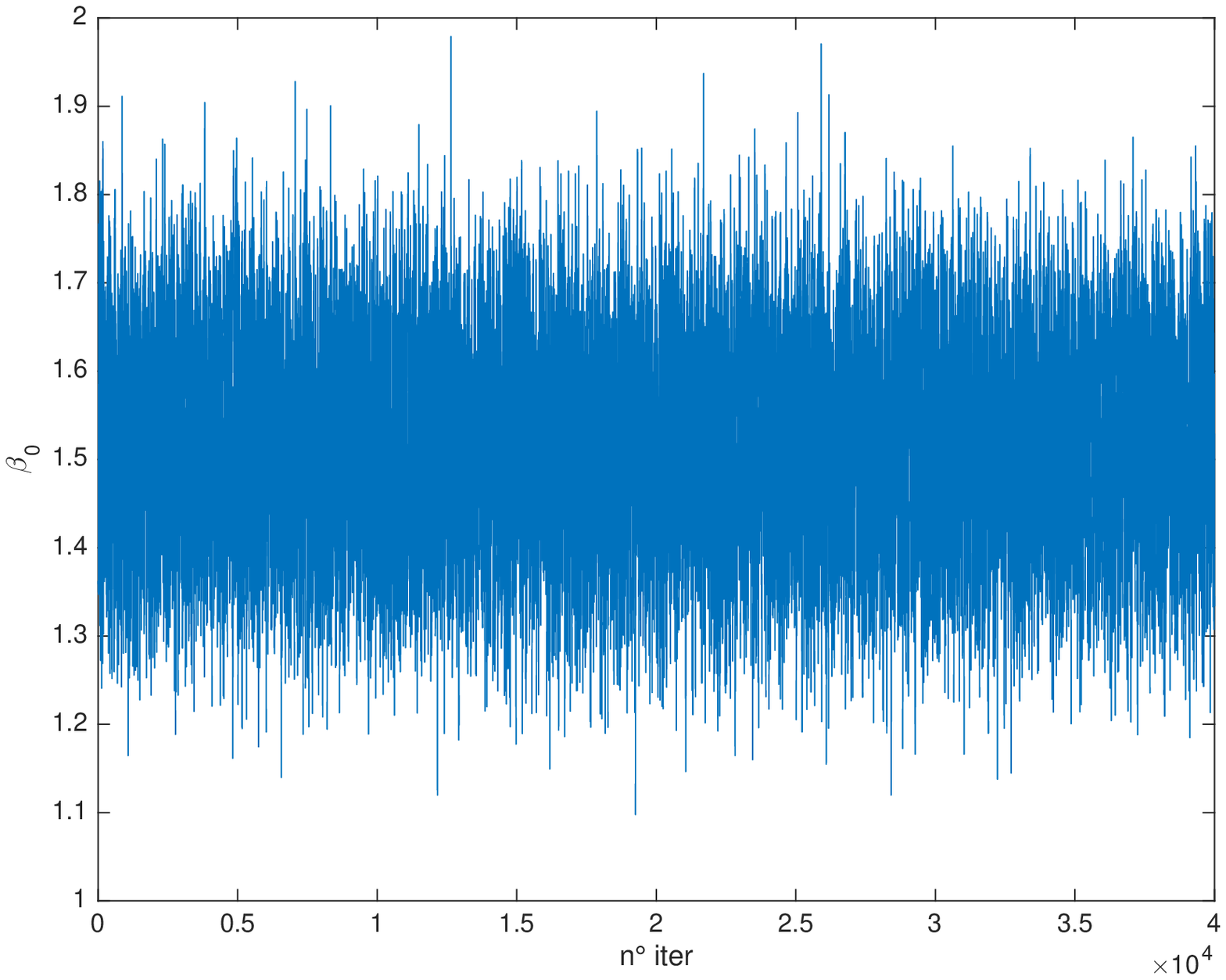}}
  	   \subfigure[]
   {\includegraphics[width=6.75cm]{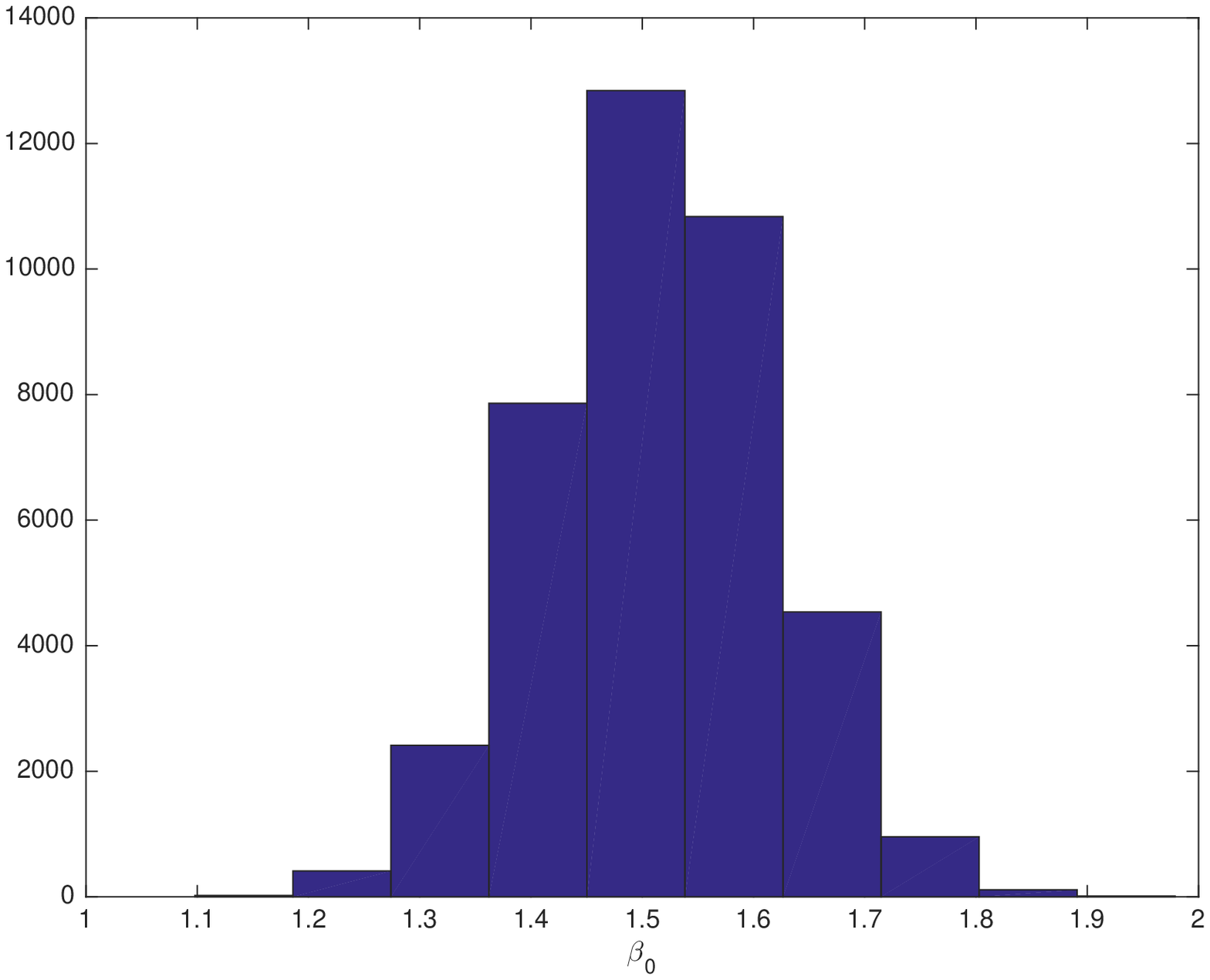}}
  	\subfigure[]
   {\includegraphics[width=6.75cm]{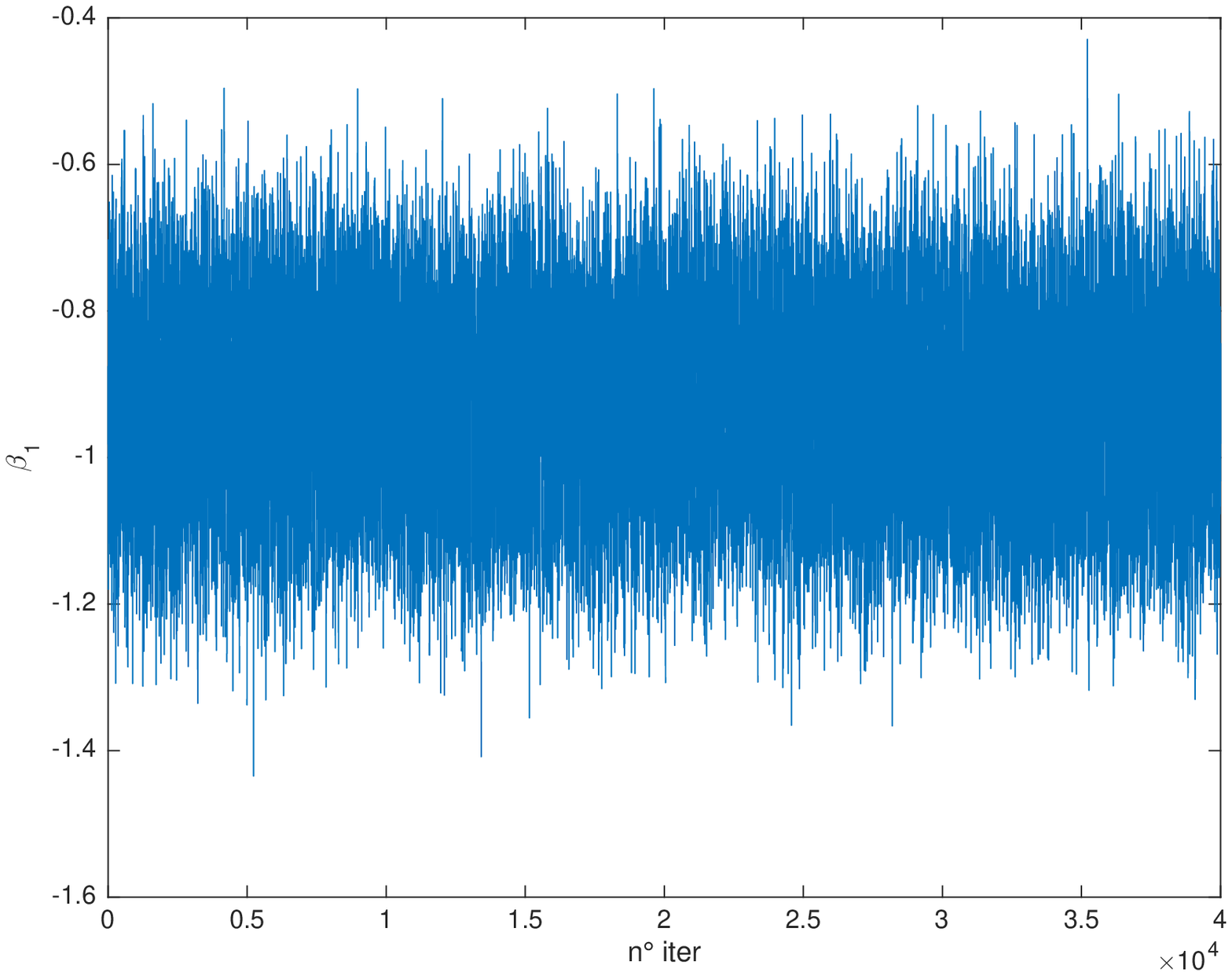}}
  \subfigure[]
   {\includegraphics[width=6.75cm]{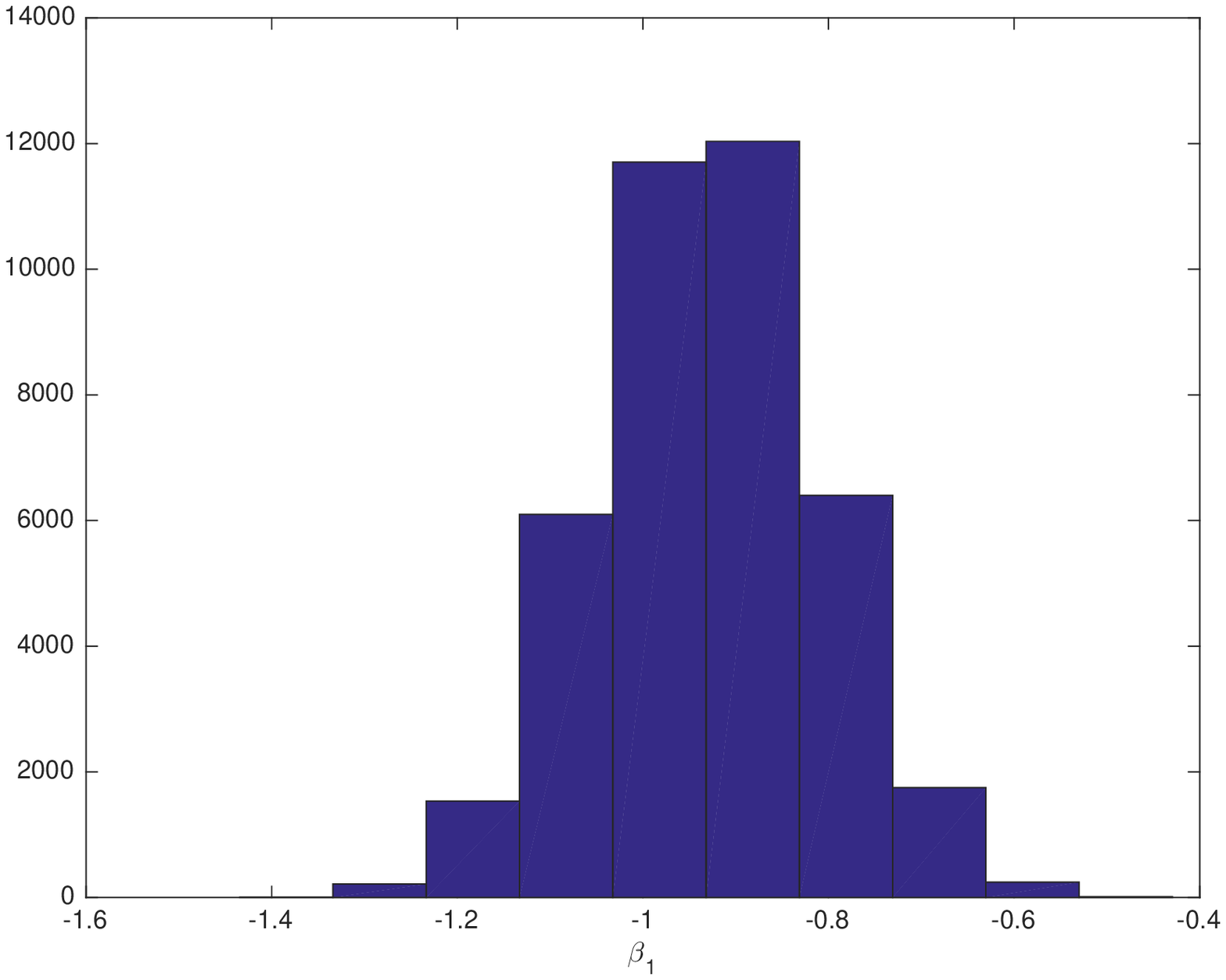}}
   	\subfigure[]
   {\includegraphics[width=6.75cm]{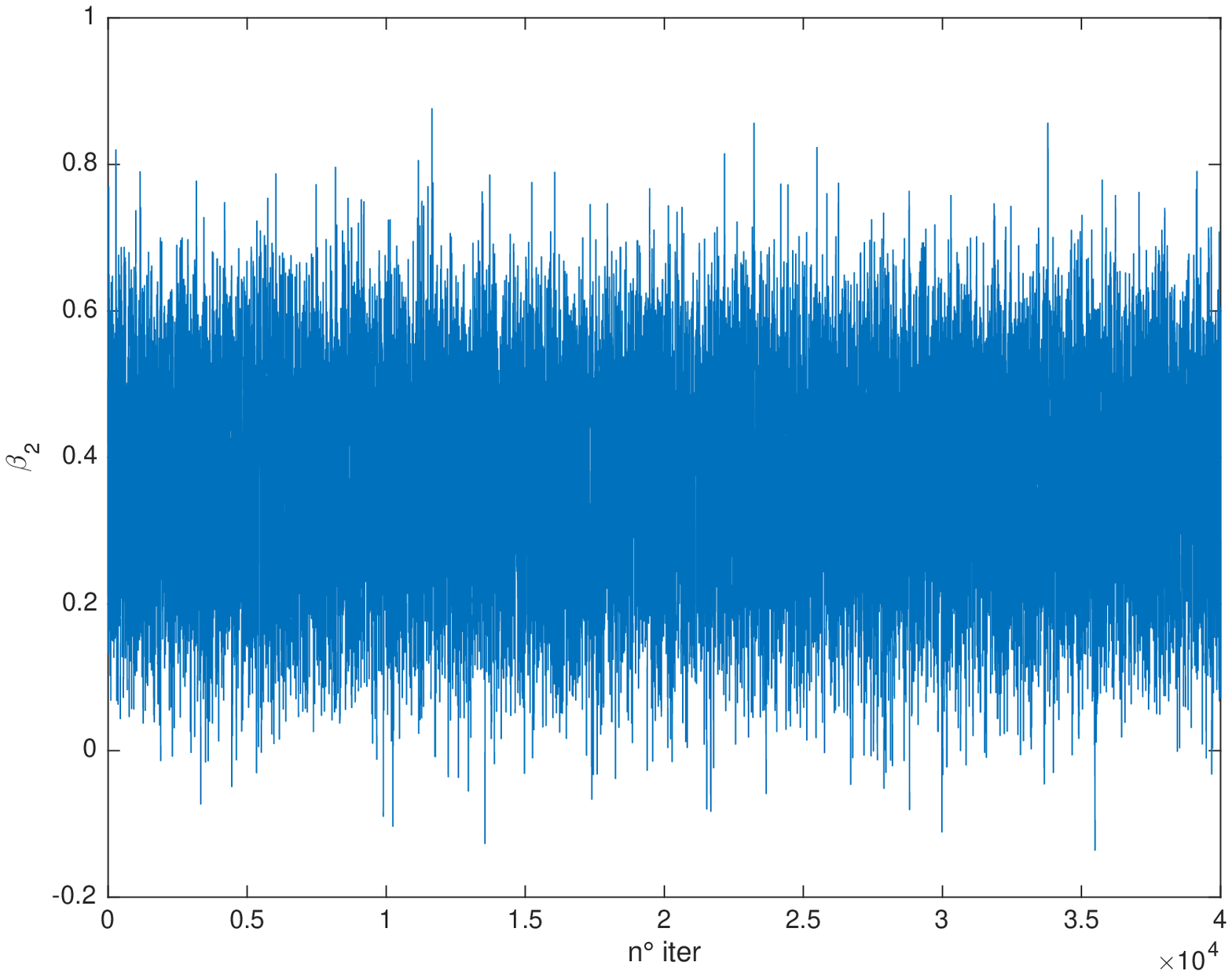}}
  \subfigure[]
   {\includegraphics[width=6.75cm]{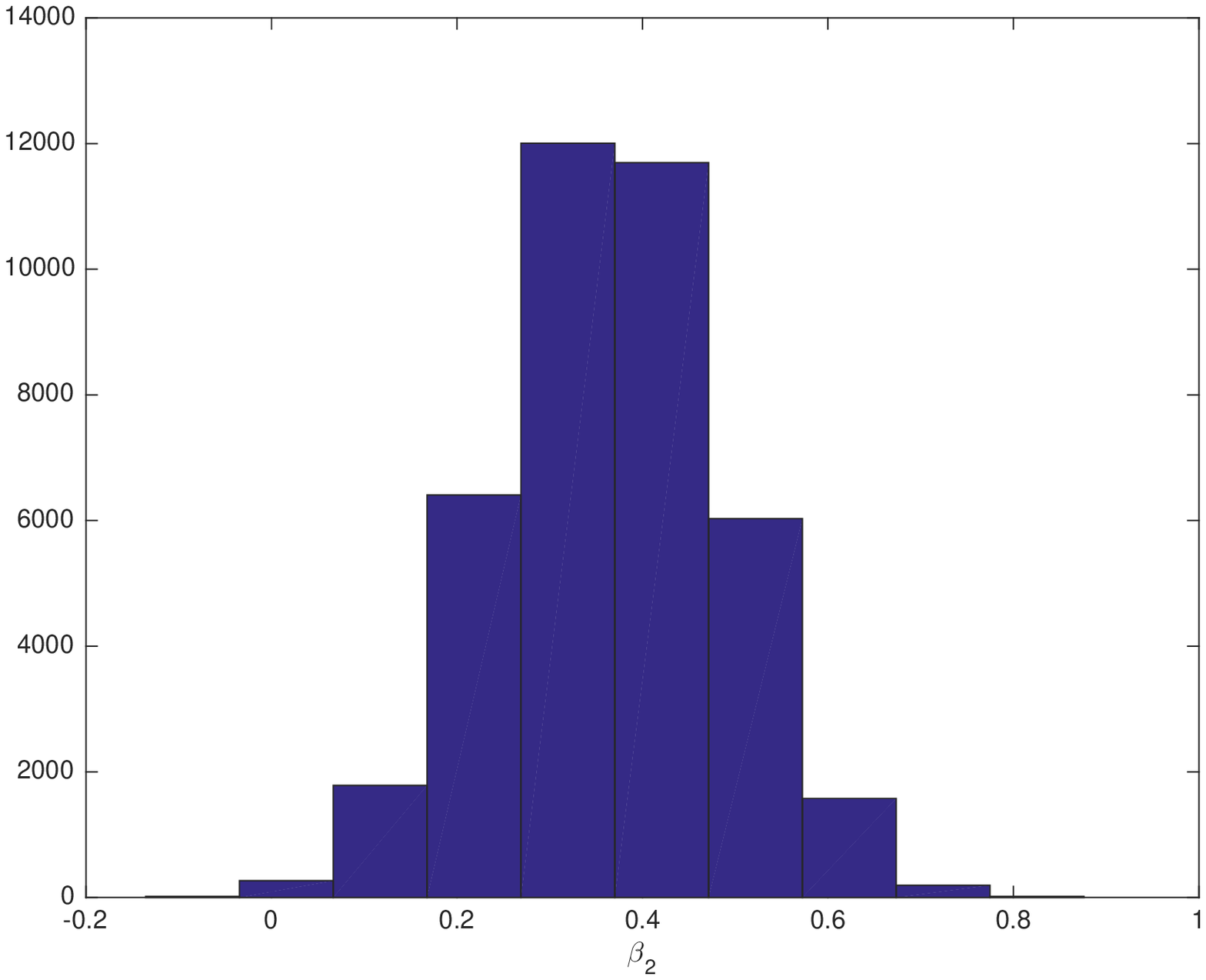}}
\caption{Posterior sample (left) and posterior histogram (right) for the simulation study of a count data regression with $\beta_0=1.5$ (top), $\beta_1=-1.0$ (middle) and $\beta_2=0.4$ (bottom) and sample size $n=300$.}
\label{FigSimCount2}
\end{figure}

As above highlighted, the procedure can be applied to multiple regressors, Figure \ref{FigSimCount2} shows the posterior samples and posterior histograms for a scenario with $\beta_0=1.5$, $\beta_1=-1.0$ and $\beta_2=0.4$. For a sample of $n=300$, and with the same setting of the Gibbs sampler used in the previous illustration, we see a good mixing of the chains as well as good inferential results. In particular, the three means for $\beta_0$, $\beta_1$ and $\beta_2$ are, respectively, 1.5, -0.9 and 0.4, with respective $95\%$ credible intervals  $(1.3,1.7)$, $(-1.2,-0.7)$ and $(0.1,0.6)$.

%%%%%%%%%%%%%%%%%%%%%%%%%%%%%%%%%%%%%%%%%%%%%%%%%%%%%%
%%    	     	Real Data Example			%%%
%%%%%%%%%%%%%%%%%%%%%%%%%%%%%%%%%%%%%%%%%%%%%%%%%%%%%%
\section{Real Data Applications: Text Analysis}
\label{Real}

%\subsection{Text Analysis}
%\label{Word}
In this section, to apply the proposed algorithm to a real-data scenario, we analyse the Yule--Simon distribution to model the word frequency in five novels: \emph{Ulysses} by James Joyce, \emph{Don Quixote} by Miguel de Cervantes, \emph{Moby Dick} by Herman Melville, \emph{War and Peace} by Leo Tolstoi and \emph{Les Miserables} by Victor Hugo. All texts are the English version present in the website of the Gutenberg Project (\href{http://www.gutenberg.org}{http://www.gutenberg.org}). We have selected the above novels as they have been analysed in \cite{Garcia11}, and we can compare our results with the author's.

The key information for each data set is $n$, the number of distinct words in the text (see Table \ref{TableRes}), and $\textbf{k}$, the frequency at which each of the words appears in the text.

\begin{figure}[h!]
\centering
	\subfigure[]
  	{\includegraphics[width=6.75cm]{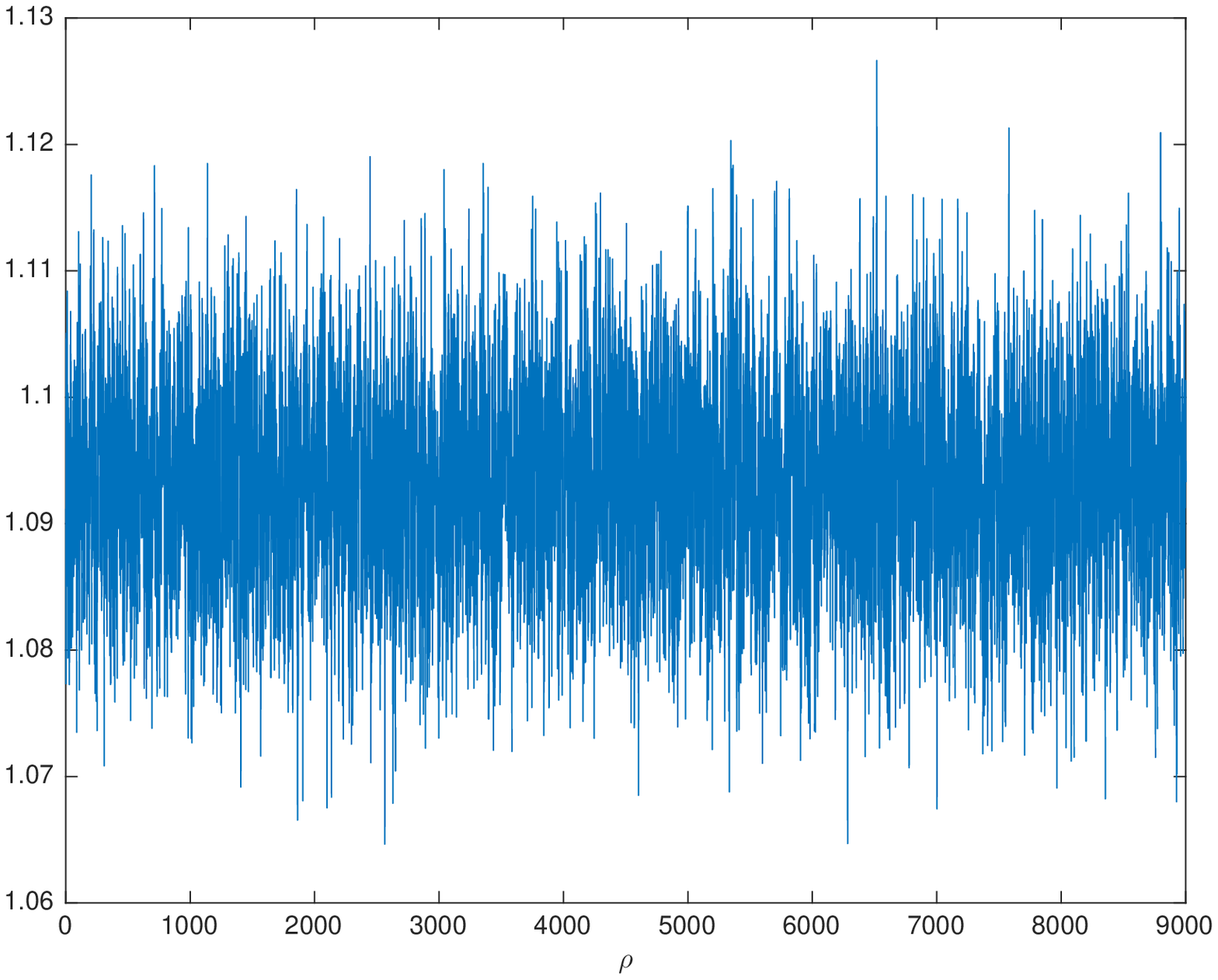}}
  	\subfigure[]
   {\includegraphics[width=6.75cm]{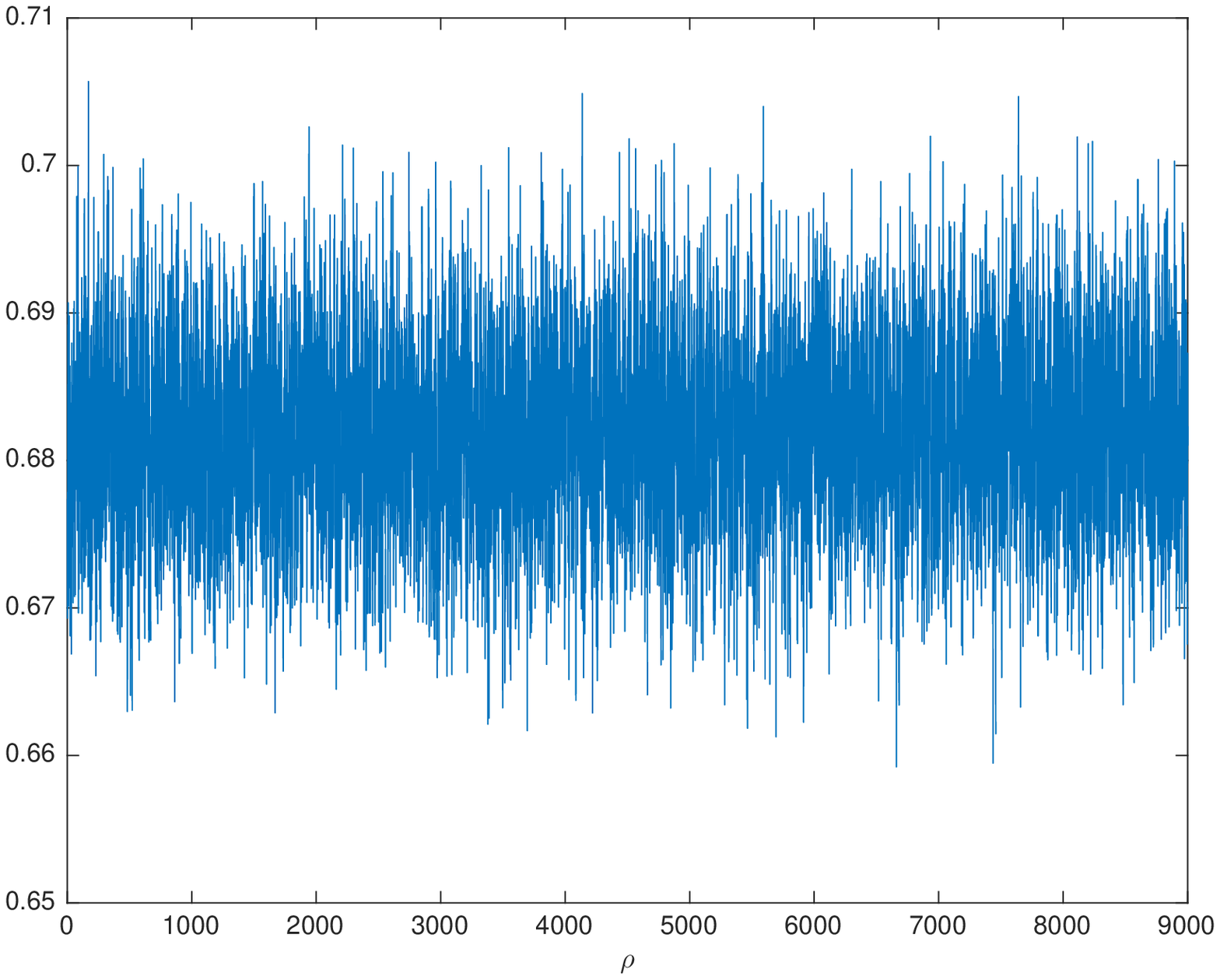}}
     \subfigure[]
   {\includegraphics[width=6.75cm]{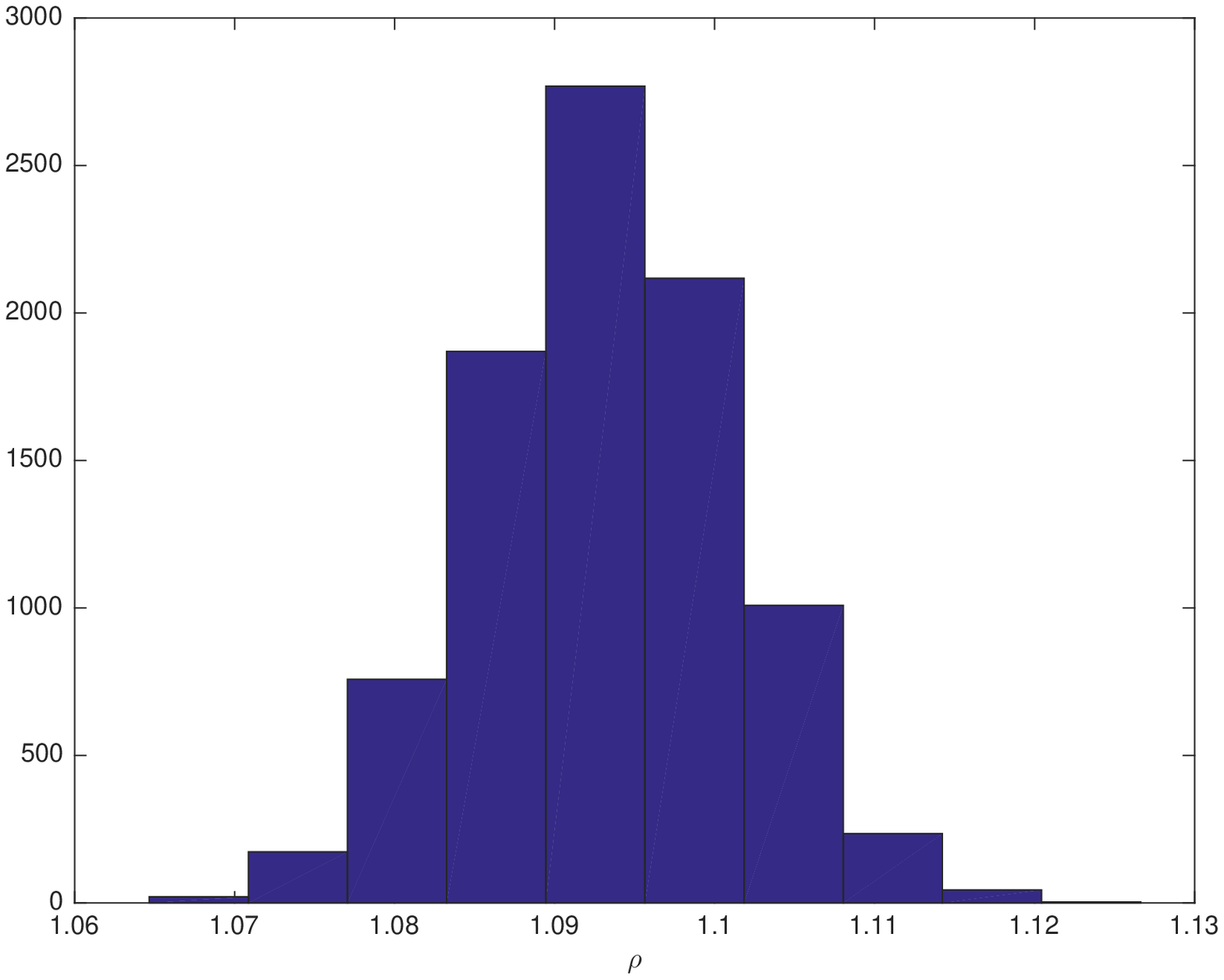}}
   \subfigure[]
   {\includegraphics[width=6.75cm]{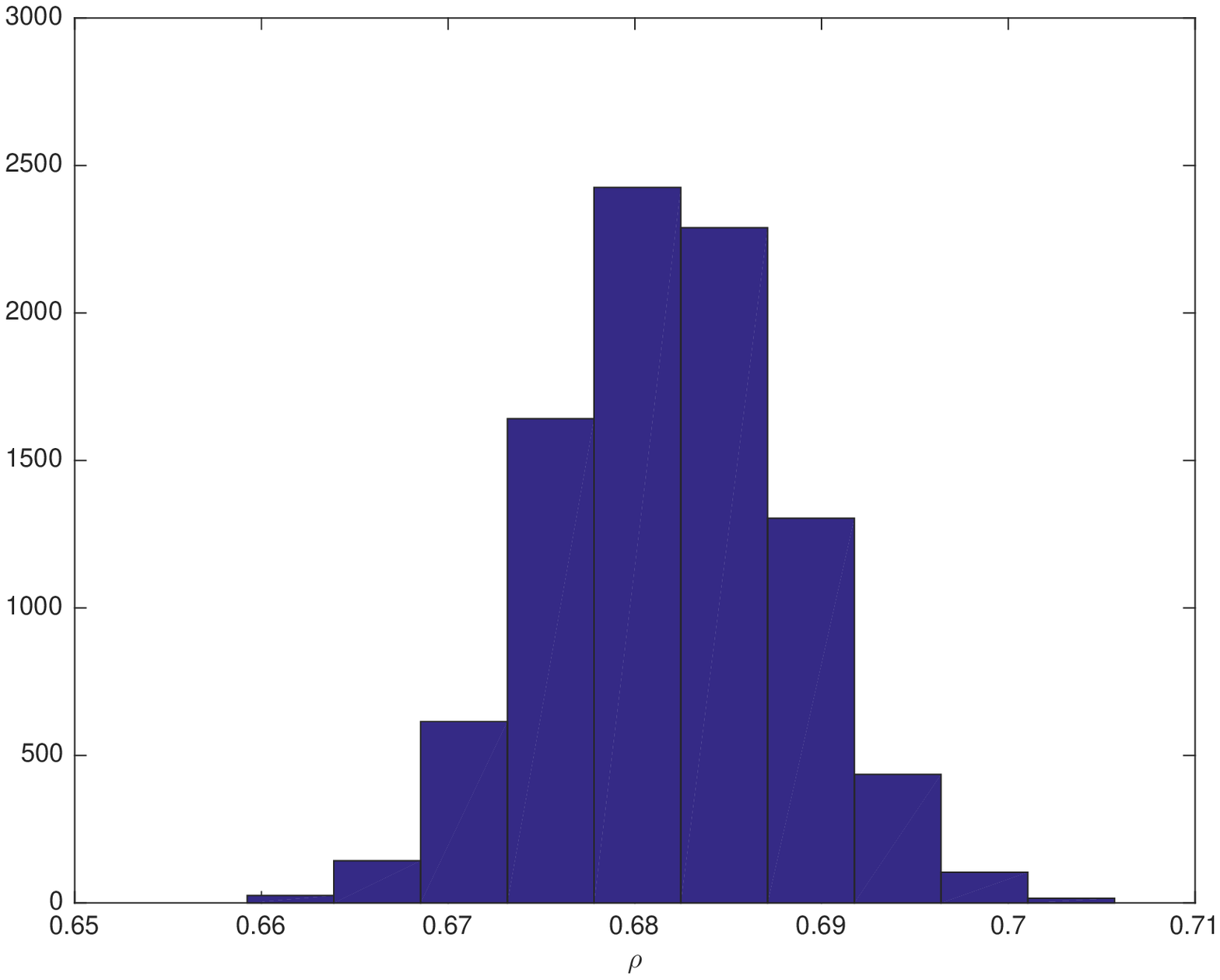}}
\caption{Posterior sample and posterior histogram for the frequency of words analysis for the Ulysses (left) and the Don Quixote (right).}
\label{FigRealRho}
\end{figure}

The inferential procedure consists in the Gibbs sampling algorithm introduced in Section \ref{BApp}. For each text, we run three chains, from different starting points, for 10,000 iterations and a burn-in period of 1,000 iterations. The convergence of the sampler has been assessed by graphical means (e.g. progressive means, Gelman and Rubin's plot) and numerical means, such as the Gelman and Rubin's convergence diagnostic and the Geweke's convergence diagnostic. The summary of a posterior for each text are shown in Table \ref{TableRes}, where we have reported the posterior mean and median, and the $95\%$ credible interval. Figure \ref{FigRealRho} shows the posterior chain and the posterior histogram for two of the analysed texts: the Ulysses and the Don Quixote.

\begin{table}[h!]
\centering
\begin{tabular}{lcccccc}
\hline
Novel & $n$ & Mean & Median  & $95\%$ C.I. & Fixed-Point Alg\\ %& Std. Dev.
\hline
Ulysses & 29,841 & 1.09 & 1.09 & (1.08,1.11) & 1.09 \\ % & 0.17 
Don Quixote & 15,180 & 0.68 & 0.68  & (0.67,0.70) & 0.68\\ % & 0.09
Moby Dick & 17,221 & 0.88 & 0.88 & (0.86,0.89) & 0.88\\
War and Peace & 18,239 & 0.63 &0.63 & (0.62,0.64) & 0.63 \\
Les Miserables & 23,248 & 0.69 & 0.69 & (0.68,0.70) & 0.69 \\
\hline
\end{tabular}
\caption{Summary statistics of the posterior distributions for the parameter $\rho$ for frequency of words compared with the fixed point algorithm.}
\label{TableRes}
\end{table}

To support our conclusions, we compare our estimation results with the ones obtained by applying the fixed-point algorithm proposed by \cite{Garcia11}. We have implemented the above algorithm on the data available to us, and the right column of Table \ref{TableRes} reports the maximum likelihood estimates for each text. First, we note that our fixed-point estimates are very similar to the results in \cite{Garcia11}, with the exception of the Don Quixote where we have used a different version of the text. Second, and most important, the mean of our posterior is virtually identical to the estimate in \cite{Garcia11}.

%%%%%%%%%%%%%%%%%%%%%%%%%%%%%%%%%%%%%%%%%%%%%%%%%%%%%%
%%    	     	Conclusions			%%%
%%%%%%%%%%%%%%%%%%%%%%%%%%%%%%%%%%%%%%%%%%%%%%%%%%%%%%
\section{Discussions}
\label{Concl}
Besides filling a gap in the Bayesian literature, the data augmentation algorithm introduced in this note, performs an efficient and fast estimation of the shape parameter of the Yule--Simon distribution.

The simulation study in Section \ref{Simu}, which discussed both a single i.i.d. sample and a count data regression sample, shows a clear out-performance of the Bayesian approach against the appropriate frequentist procedures. This is particularly true for relatively small sample sizes, rendering the Bayesian inference for the Yule--Simon distribution attractive to practitioners.

The real data example illustrated in Section \ref{Real} shows the soundness of the approach when true observations are considered. Furthermore, the obtained results are cross-validated by the equivalence of the results in \cite{Garcia11} obtained through a frequentist procedure.

%%%%%%%%%%%%%%%%%%%%%%%%%%%%%%%%%%%%%%%%%%%%%%%%%%%%%%
%%    	     	Acknowledgements					%%%
%%%%%%%%%%%%%%%%%%%%%%%%%%%%%%%%%%%%%%%%%%%%%%%%%%%%%%
\section*{Acknowledgements}
The authors are thankful to the Associate Editor and the anonymous reviewers for their useful
comments which significantly improved the quality of the paper. Fabrizio Leisen was supported by the European Community's Seventh Framework
Programme [FP7/2007-2013] under grant agreement no: 630677.
\vspace*{-8pt}

%%%%%%%%%%%%%%%%%%%%%%%%%%%%%%%%%%%%%%%%%%%%%%%%%%%%%%
%%          Bibliography 		%%%
%%%%%%%%%%%%%%%%%%%%%%%%%%%%%%%%%%%%%%%%%%%%%%%%%%%%%%
%\bibliographystyle{apalike}
\bibliographystyle{elsarticle-harv} 
\bibliography{YuSimoBiblio}

\end{document}